\documentclass{jnmp}

\begin{document}
\setcounter{page}{217}
\renewcommand{\evenhead}{S~Yu~Sakovich}
\renewcommand{\oddhead}{A System of Four ODEs: The Singularity Analysis}

\thispagestyle{empty}

%\FistPageHead{2}{\pageref{sakovich-firstpage}
%--\pageref{sakovich-lastpage}}{Letter}

\FirstPageHead{8}{2}{2001}
{\pageref{sakovich-firstpage}--\pageref{sakovich-lastpage}}{Letter}

\copyrightnote{2001}{S~Yu~Sakovich}

\Name{A System of Four ODEs:
The Singularity Analysis}\label{sakovich-firstpage}

\Author{S~Yu~SAKOVICH}

\Address{Institute of Physics, National Academy of Sciences, 220072 Minsk, Belarus\\
E-mail: saks@pisem.net}

\Date{Received November 23, 2000; Revised November 29, 2000; Accepted  January 10, 2001}

\begin{abstract}
\noindent
The singularity analysis is carried out for a system of four first-order
quadratic ODEs with a parameter, which was proposed recently by Golubchik and
Sokolov. A transformation of dependent variables is revealed by the analysis,
after which the transformed system possesses the Painlev\'{e} property and
does not contain the parameter.
\end{abstract}

\noindent
Recently, Golubchik and Sokolov \cite{sakovich:GS} proposed the following system of
four first-order quadratic ODEs:
\begin{equation}\arraycolsep=0em
\begin{array}{l}
p_{t}=p^{2}-pr-qs,
\vspace{1mm}\\
q_{t}=apq+(a-2)rq,
\vspace{1mm}\\
r_{t}=r^{2}-pr-qs,
\vspace{1mm}\\
s_{t}=(1-a)ps+(3-a)rs,
\end{array}
\label{sakovich:gss}
\end{equation}
where $a$ is a parameter. It was pointed out in~\cite{sakovich:GS} that the system
(\ref{sakovich:gss}), though integrable by quadratures, probably does not pass the
Painlev\'{e}--Kovalevskaya test for generic~$a$. More recently, Leach, Cotsakis
and Flessas~\cite{sakovich:LCF} drew a conclusion that the system (\ref{sakovich:gss}) does not
possess the Painlev\'{e} property for any value of the parameter~$a$.

In the present note, we give our version of the singularity analysis of the
system~(\ref{sakovich:gss}). The analysis reveals a very simple transformation of the
variables $q$ and $s$, after which the transformed system possesses the
Painlev\'{e} property and does not contain the parameter~$a$.

Let us carry out the singularity analysis for the system~(\ref{sakovich:gss}),
following the Ablowitz--Ramani--Segur algorithm~\cite{sakovich:ARS} (see also~\cite{sakovich:RGB}).
Substituting into (\ref{sakovich:gss}) the expansions of $p$, $q$, $r$, $s$ near
$\phi(t)=0 $, $\phi_{t}=1$,
\begin{equation}\arraycolsep=0em
\begin{array}{ll}
p=p_{0}\phi^{\alpha}+\cdots+p_{n}\phi^{n+\alpha}+\cdots, \qquad & q=q_{0}\phi^{\beta
}+\cdots+q_{n}\phi^{n+\beta}+\cdots,
\vspace{1mm}\\
r=r_{0}\phi^{\gamma}+\cdots+r_{n}\phi^{n+\gamma}+\cdots,
\qquad & s=s_{0}\phi
^{\delta}+\cdots+s_{n}\phi^{n+\delta}+\cdots,
\end{array}
\label{sakovich:exp}
\end{equation}
we find the following three branches (i.e. the admissible choices of
$\alpha$, $\beta$, $\gamma$,
$\delta$ and $p_{0}$, $q_{0}$, $r_{0}$, $s_{0}$ with the
corresponding positions $n$ of resonances), besides the evident branch
governed by the Cauchy theorem ($\alpha=\beta=\gamma=\delta=0$, $\forall\;
p_{0},\; q_{0},\; r_{0},\; s_{0}$):
\begin{equation}\arraycolsep=0em
\begin{array}{l}
\alpha=-1,\qquad \beta=-a,\qquad \gamma=0,\qquad \delta=a-1,
\vspace{1mm}\\
p_{0}=-1,\qquad r_{0}=q_{0}s_{0},\quad \forall \; q_{0}, \; s_{0},
\qquad (n+1)n^{2}(n-1)=0;
\end{array}
\label{sakovich:I}
\end{equation}
\begin{equation}\arraycolsep=0em
\begin{array}{l}
\alpha=0,\qquad \beta=2-a,\qquad \gamma=-1,\qquad \delta=a-3,
\vspace{1mm}\\
r_{0}=-1,\qquad p_{0}=q_{0}s_{0},\quad \forall \; q_{0},\; s_{0},\qquad
(n+1)n^{2}(n-1)=0;
\end{array}
\label{sakovich:II}
\end{equation}\arraycolsep=0em
\begin{equation}
\begin{array}{l}
\alpha=-1,\qquad \beta=2-2a,\qquad \gamma=-1,\qquad \delta=2a-4,
\vspace{1mm}\\
p_{0}=r_{0}=-1,\qquad q_{0}s_{0}=-1,\quad \forall \; q_{0}\ \ \mathrm{xor}\  \ \forall \; s_{0},
\qquad (n+1)^{2}n(n-2)=0.
\end{array}
\label{sakovich:III}
\end{equation}

We see from (\ref{sakovich:I}), (\ref{sakovich:II}), (\ref{sakovich:III})
that the system (\ref{sakovich:gss}) may
possess the Painlev\'{e} property only if the parameter~$a$ is integer. But
the positions of resonances are integer and independent of~$a$ in all the
branches, and this suggests that the expansions~(\ref{sakovich:exp}) do not contain
terms with noninteger $n$. Moreover, the expansions (\ref{sakovich:exp}) are free from
logarithmic terms, as we can prove by checking the consistency of recursion
relations for $p_{n}$, $q_{n}$, $r_{n}$, $s_{n}$, $n=0,1,\ldots$, obtainable from
(\ref{sakovich:gss}). We have the following:
\[
p_{1}=0,\qquad q_{1}=(a-2)q_{0}^{2}s_{0},\qquad s_{1}=(3-a)q_{0}s_{0}^{2},\quad \forall\;
r_{1},
\]
for (\ref{sakovich:I});
\[
r_{1}=0,\qquad q_{1}=aq_{0}^{2}s_{0},\qquad s_{1}=(1-a)q_{0}s_{0}^{2},\quad \forall \; p_{1},
\]
for (\ref{sakovich:II});
\[\arraycolsep=0em
\begin{array}{l}
p_{1}=q_{1}=r_{1}=s_{1}=0,\qquad p_{2}=r_{2}=-s_{0}q_{2}-q_{0}s_{2},
\vspace{1mm}\\
(a-2)s_{0}q_{2}+(a-1)q_{0}s_{2}=0,\quad  \forall \; q_{2}\ \ \mathrm{xor}\  \ \forall\;
s_{2},
\end{array}
\]
for (\ref{sakovich:III}). Consequently, in all the branches, solutions of (\ref{sakovich:gss})
are represented by the expansions
\[
\displaystyle p=\phi^{\alpha}\sum_{n=0}^{\infty}p_{n}\phi^{n},
\qquad   \displaystyle q=\phi^{\beta
}\sum_{n=0}^{\infty}q_{n}\phi^{n},
\qquad
\displaystyle r=\phi^{\gamma}\sum_{n=0}^{\infty}r_{n}\phi^{n}, \qquad
 \displaystyle s=\phi^{\delta}\sum
_{n=0}^{\infty}s_{n}\phi^{n},
\]
where $\alpha,\gamma=-1,0$, and the functions $\beta(a)$ and $\delta(a)$ vary
from branch to branch. Therefore we can hope to improve the analytic
properties of the system~(\ref{sakovich:gss}) by an appropriate transformation of the
dependent variables $q$ and $s$.

Let us consider a variable $z(t)$,
\begin{equation}
z=q^{x}s^{y},\label{sakovich:z}
\end{equation}
where $x$ and $y$ are constants, and study its dominant behavior $z=z_{0}
\phi^{\epsilon}+\cdots$ near $\phi=0$ in the branches (\ref{sakovich:I}), (\ref{sakovich:II}),
(\ref{sakovich:III}). We have $\epsilon=i$,
\begin{equation}
i=-ax+(a-1)y,\label{sakovich:i}
\end{equation}
in the branch (\ref{sakovich:I}); $\epsilon=j$,
\begin{equation}
j=(2-a)x+(a-3)y,\label{sakovich:j}
\end{equation}
in the branch (\ref{sakovich:II}); $\epsilon=k$,
\begin{equation}
k=(2-2a)x+(2a-4)y,\label{sakovich:k}
\end{equation}
in the branch (\ref{sakovich:III}). Since $k=i+j$ due to (\ref{sakovich:i}), (\ref{sakovich:j}),
(\ref{sakovich:k}), $k$ will be integer for any integer~$i$ and~$j$. And, for any
integer~$i$ and $j$, the variable $z$ (\ref{sakovich:z})
will possess a good dominant
behavior in each of the branches (\ref{sakovich:I}),
(\ref{sakovich:II}), (\ref{sakovich:III}), if we set
\begin{equation}
x=\frac{1}{2}(a-3)i-\frac{1}{2}(a-1)j,\qquad
y=\frac{1}{2}(a-2)i-\frac{1}{2}aj
\label{sakovich:xy}
\end{equation}
due to (\ref{sakovich:i}), (\ref{sakovich:j}).

Which choice of the integers $i$ and $j$ to prefer? According to (\ref{sakovich:gss})
and (\ref{sakovich:z}),
\[
z_{t}=(-ip-jr)z,
\]
and it seems natural to choose $i=-1$, $j=0$ or $i=0$, $j=-1$. Denoting
$z|_{i=-1,j=0}$ as $u$, and $z|_{i=0,j=-1}$ as $v$, we find from (\ref{sakovich:z}),
(\ref{sakovich:xy}) that
\[
u=q^{\frac{1}{2}(3-a)}s^{\frac{1}{2}(2-a)},\qquad  v=q^{\frac{1}{2}(a-1)}s^{\frac
{1}{2}a}.
\]
In the new variables $p$, $r$, $u$, $v$, the system (\ref{sakovich:gss}) changes into
\begin{equation}\arraycolsep=0em
\begin{array}{l}
p_{t}=p^{2}-pr-uv,
\vspace{1mm}\\
r_{t}=r^{2}-pr-uv,
\vspace{1mm}\\
u_{t}=pu,
\vspace{1mm}\\
v_{t}=rv.
\end{array}
\label{sakovich:new}
\end{equation}
The system (\ref{sakovich:new}) possesses the Painlev\'{e} property and does not
contain the parameter $a$. Also, the three constants of motion of the system
(\ref{sakovich:gss}), given in \cite{sakovich:GS} in the variables $p$,
$q$, $r$, $s$, become more simple
in the new variables $p$, $r$, $u$, $v$:
\[
c_{1}=pr+uv,\qquad c_{2}=\frac{p-r}{uv},\qquad c_{3}=\frac{r^{2}-pr-uv}{v^{2}}.
\]
The constant of motion $c_{4}=(p^{2}-pr-uv)/u^{2}$ simply follows from $c_{3}$
via the evident symmetry $p\leftrightarrow r$, $u\leftrightarrow v$ of the
system (\ref{sakovich:new}), and one can use any three of $c_{1}$,
$c_{2}$, $c_{3}$, $c_{4}$ as mutually independent.

Obviously, the analytic properties of the Golubchik--Sokolov system~(\ref{sakovich:gss})
do not contradict to the empirically well known interrelation between the
Painlev\'{e} property and the integrability of ODEs and PDEs. This system is
similar to the sine-Gordon equation, which also does not possess the
Painlev\'{e} property until a simple transformation is made~\cite{sakovich:W}.

\label{sakovich-lastpage}
\end{document}